# Remarks On The Standard Hylleraas-Undheim And MacDonald Computation Of Excited States

Naoum C. Bacalis

*Theoretical and Physical Chemistry Institute, National Hellenic Research Foundation, Vasileos Constantinou 48, GR-116 35 Athens, Greece*

For the computation of excited states, the standard solutions of the Schroedinger equation, using higher roots of a secular equation in a finite N-dimensional function space, by the Hylleraas-Undheim and MacDonald theorem, have several restrictions, which render them of lower quality, relative to the lowest root, if the latter is good enough. These deficiencies are reported, that prevent from comparisons with accurate experiments.

## I. INTRODUCTION AND NOTATION

The standard solutions of the Schroedinger equation for bound excited states are obtained as higher roots of a secular equation according to the Hylleraas-Undheim and MacDonald (HUM) theorem, [1] which warrants that they are orthogonal to the lowest root and have higher energy than the exact excited state. Although the projection of the lowest HUM root on the exact ground state $|0\rangle$ obeys $|\langle 0|0^{HUM}\rangle|^2 \leq 1$ (all wave functions are assumed normalized and real), so that, by the Eckart variational (energy minimization) theorem, [2] it is allowed to approach 1 at will, however, the 2nd HUM root is restricted to have $|\langle 1|1^{HUM}\rangle|^2 < 1 - |\langle 1|0^{HUM}\rangle|^2$, so that, it is not allowed to approach 1 at will, unless $|\langle 1|0^{HUM}\rangle| = 0$ (very unlikely in an N-dimensional finite function space). In addition, for any $|0^{HUM}\rangle$ there are at least two wave functions, $|f^I\rangle$, $|f^{II}\rangle$, both orthogonal to $|0^{HUM}\rangle$, which are not very good approximations of the exact $|1\rangle$, but have exactly its energy, $E[f^I] = E_1 = E[f^{II}]$, whereas $E[1^{HUM}] > E_1$. Therefore, there is a function $|f^{I\_best}\rangle$, such that $|\langle 1|1^{HUM}\rangle|^2 < |\langle 1|f^{I\_best}\rangle|^2 < 1 - |\langle 1|0^{HUM}\rangle|^2$. [The last inequality is based on the fact that among all functions $|\phi_1\rangle$ orthogonal to $|0^{HUM}\rangle$, the closest to $|1\rangle$, (labeled $|\phi_1^+\rangle$) has lower energy: $E[\phi_1^+] < E_1$]. This restriction of $|1^{HUM}\rangle$ prevents from explaining accurate experiments. (In order to avoid this deficiency, and because the excited states cannot be obtained variationally by minimization of the energy, since they are saddle points in the Hamiltonian eigenfunction Hilbert space, variational functionals, for a non-degenerate Hamiltonian, have been reported (cf. arXiv:0801.3673), which have local minimum at the excited states.)

## II. REMARKS ON "HUM" HIGHER ROOTS

An analysis is presented, supposing (an actually reasonable assumption) that $s \equiv \langle 1|0^{HUM}\rangle \neq 0$ and that $\langle 0|0^{HUM}\rangle$ is large enough, with energy $E[0^{HUM}] \equiv \langle 0^{HUM}|H|0^{HUM}\rangle \approx E_0 < E_1$, while $\langle 1^{HUM}|0^{HUM}\rangle = 0$. Then, as shown below, $|1^{HUM}\rangle$ is restricted, by HUM theorem, to have

$$\langle 1|1^{HUM}\rangle^2 < \left(1 - \langle 1|0^{HUM}\rangle^2\right) \bigg/ \left(1 + \frac{E_1 - E[0^{HUM}]}{-E_1}\langle 1|0^{HUM}\rangle^2 \bigg/ \left(1 - \langle 1|0^{HUM}\rangle^2\right)\right) < \left(1 - \langle 1|0^{HUM}\rangle^2\right) \qquad (1)$$

If $\langle 0|0^{HUM}\rangle$ is large enough, then it will be shown that $\langle 1|1^{HUM}\rangle^2 < \langle 0|0^{HUM}\rangle^2$, so that, $|1^{HUM}\rangle$ has lower quality than $|0^{HUM}\rangle$.

The proof is given in three steps. Given $|0^{HUM}\rangle$: (i) $|1^{HUM}\rangle$ cannot approach $|1\rangle$ due to orthogonality to $|0^{HUM}\rangle$. (ii) $|1^{HUM}\rangle$ cannot approach $|\phi_1^+\rangle$, the "best" orthogonal to $|0^{HUM}\rangle$, because $E[\phi_1^+] < E_1 < E[1^{HUM}]$. (iii) $|1^{HUM}\rangle$ cannot approach $|f^{I\_best}\rangle$, for which $E[\phi_1^+] < E[f^{I\_best}] \equiv E_1 < E[1^{HUM}]$, and which is orthogonal to $|0^{HUM}\rangle$ but is "worse" than $|\phi_1^+>$. All these functions, $|\phi_1^+>$, $|f^{I\_best}>$, $|1^{HUM}>$, converge to $|1>$ when $|0^{HUM}>$ tends to $|0>$, but they spread away, as described below, when $|0^{HUM}>$ departs from $|0>$.

### (i) $|1^{HUM}>$ Cannot Approach $|1>$ Due To Orthogonality To $|0^{HUM}>$

If $s \equiv <1|0^{HUM}> \neq 0$, then among all wave functions $|\phi_1>$, that are orthogonal to $|0^{HUM}>$, the Gram-Schmidt orthogonal to $|0^{HUM}>$ on the subspace of $\{|1>, |0^{HUM}>\}$, which has the largest component on $|1>$, is

$$|\phi_1^+> \equiv (|1> - s\,|0^{HUM}>)/(1-s^2)^{1/2}, \tag{2}$$

Therefore, $|1^{HUM}>$ cannot approach $|1>$ more than $|\phi_1^+>$ : $<1|1^{HUM}>^2 \leq <1|\phi_1^+>^2$. We shall see that only "<" holds.

[Incidentally, it is parenthetically noted that, in an actual calculation, if $s = <1|\phi_0> \neq 0$, then the strict Eckart lower bound of an approximant ground state is, in practice, too low: If $s \neq 0$, then

$$E[\phi_0] = E_0 + (E_1 - E_0)<1|\phi_0>^2 + (E_2 - E_0)<2|\phi_0>^2 + \ldots \geq E_0 + (E_1 - E_0)\,s^2,$$

the actual lowest energy that a calculation can report, because, in practice, always $s \neq 0$.]

### (ii) $|1^{HUM}>$ Cannot Approach $|\phi_1^+>$ Because $E[\phi_1^+] < E_1 \leq E[1^{HUM}]$

By HUM theorem, if $|0^{HUM}> \neq |0>$, then $E[1^{HUM}] > E_1$. However, the energy of $|\phi_1^+>$ is

$$E[\phi_1^+] = <\phi_1^+|H|\phi_1^+> = E_1 - (E_1 - E[0^{HUM}])\,s/(1-s^2) < \text{(less than) } E_1, \tag{3}$$

Therefore, if $s \equiv <1|0^{HUM}> \neq 0$, then $|1^{HUM}>$ cannot approach $|\phi_1^+>$ itself: $<1|1^{HUM}>^2 << <1|\phi_1^+>^2$ (strict inequality).

Since $E[\phi_1^+] < E_1 < E[1^{HUM}]$, there are at least two wave functions orthogonal to $|0^{HUM}>$, with the energy of $|1>$: In principle, these are obtained as follows: On the subspace of $\{|1^{HUM}>, |\phi_1^+>\}$ (which is orthogonal to $|0^{HUM}>$), use $|\phi_1^+>$ and its Gram-Schmidt orthogonal, $|f_1> [\equiv (|1^{HUM}> - |\phi_1^+><\phi_1^+|1^{HUM}>)/(1 - <\phi_1^+|1^{HUM}>^2)^{1/2}]$, to compute the two Hamiltonian eigenfunctions, $|p>$, $|m>$, (where $p$ corresponds to + and $m$ to −):

$$|p>,|m> = |\phi_1^+>\frac{\left\langle\phi_1^+|H|f_1\right\rangle}{\sqrt{\rho_\pm\varepsilon_\pm}} + |f_1>\frac{\varepsilon_\pm}{\sqrt{\rho_\pm\varepsilon_\pm}}, \qquad E[p,m] = E_\pm = E[\phi_1^+] + \varepsilon_\pm, \tag{4}$$

where $\rho_\pm = \pm\sqrt{(E[f_1]-E[\phi_1^+])^2 + 4\left\langle\phi_1^+|H|f_1\right\rangle^2}$, and $\varepsilon_\pm = \frac{1}{2}(\rho_\pm + E[f_1] - E[\phi_1^+])$

They have wider energies, i.e: $E[m] \leq E[\phi_1^+] < E_1 < E[1^{HUM}] \leq E[p]$, because the Hamiltonian opens the energy gap of $|\phi_1^+>$ and $|1^{HUM}>$. [Therefore, $E[\phi_1^+]$ is not at the minimum, and the lowest of $E[m]$, obtained by minimizing $E[\phi_1]$ ($|\phi_1>$ orthogonal to $|0^{HUM}>$) leads to a function which is orthogonal to $|0^{HUM}>$, far from $|1>$, and lying below $E_1$]. Then, the two functions

$$|f^{I,II}> \equiv |p>\sqrt{\frac{E_1 - E[m]}{E[p] - E[m]}} \pm |m>\sqrt{\frac{E[p] - E_1}{E[p] - E[m]}}\,; \qquad E[f^{I,II}] \equiv E_1, \tag{5}$$

have $E[f^{I,II}] \equiv E_1$, and are orthogonal to $|0^{HUM}>$. Similarly, by using, instead of $|1^{HUM}>$, any function $|\phi_1>$ orthogonal to $|0^{HUM}>$, belonging to subspaces other than $\{|1^{HUM}>, |\phi_1^+>\}$ one could obtain functions like $|f^{I,II}>$, orthogonal to $|0^{HUM}>$, with the energy of $E_1$. One of these, $|f^{I\_best}>$, has the largest possible projection on $|1>$, and is obtained by some function $|\phi>$, through $|f_1>$, its Gram-Schmidt orthogonal to both $|0^{HUM}>$ and $|\phi_1^+>$, which we are looking for. In the following it is shown that for bound states below continuum (the usual cases of atoms and molecules) this largest possible projection $<1|f^{I\_best}>$ is smaller than $|<1|\phi_1^+>|$, by $(E_1/E[\phi_1^+])^{1/2}$.

The proof is as follows: Consider any $|\phi>$, orthogonalize it to $|0^{HUM}>$ to obtain $|\phi_1>$, and repeat the above procedure, replacing $|1^{HUM}>$ by $|\phi_1>$. On the subspace of $\{|\phi_1>, |\phi_1^+>\}$ compute $|f_1>$ orthogonal to $|\phi_1^+>$ (or to $|\phi_1>$), compute the two eigenfunctions $|p>$, $|m>$ and the corresponding $|f^{I,II}>$ by Eq. 5. For fixed $|0^{HUM}> \neq |0>$, their projection on $|1\rangle$ depend on $E[f_1]$ and on $\langle\phi_1^+|H|f_1\rangle$ which, in varying $|\phi\rangle$, take on various values between $E_0(<0)$ and 0, and between $E_0$ and $-E_0$ respectively. Their projection has the form:

$$\langle 1|f^{I,II}\rangle = \frac{1}{2}\left(\sqrt{1+\frac{1}{\sqrt{y}}}\sqrt{1+\frac{1}{\sqrt{y}}-\frac{2\delta}{\Delta\sqrt{y}}} \pm \sqrt{1-\frac{1}{\sqrt{y}}}\sqrt{1-\frac{1}{\sqrt{y}}+\frac{2\delta}{\Delta\sqrt{y}}}\right)\langle 1|\phi_1^+\rangle , \qquad (6)$$

where $y = \sqrt{1+4\langle\phi_1^+|H|f_1\rangle^2/\Delta^2}$ , $\Delta = E[f_1] - E[\phi_1^+]$, $\delta = E_1 - E[\phi_1^+]$.

The (+) case, $\langle 1|f^{II}\rangle$, has maximum when $y \to 0$ ($\Delta$, or $E[f_1]$, $\to \infty$ unless $\langle\phi_1^+|H|f_1\rangle = 0$), which equals $\langle 1|\phi_1^+\rangle$, (corresponding to $|\phi_1^+\rangle$ with vanishing contribution of highly excited $|f_1\rangle$).

The ($-$) case, $\langle 1|f^I\rangle$, has maximum when $y = 1$ and $\Delta$, or $E[f_1]$, $\to \infty$ . For each $E[f_1] < \infty$ , $\langle 1|f^I\rangle$ has a conditional maximum, when $y = 1$, or $\langle\phi_1^+|H|f_1\rangle = 0$, i.e. when $|f_1\rangle$, $|\phi_1^+\rangle$ become eigenvectors (on their planar subspace). The maximum equals $(1-\delta/\Delta)^{1/2}\langle 1|\phi_1^+\rangle$ and increases with increasing $E[f_1]$.

[Incidentally, $\langle 1|f^I\rangle = 0$ when $E[f_1] = E[\phi_1^+]$. This means that there are infinitely many functions $|f^I\rangle$ with the energy of $E_1$, which, however, are orthogonal to both $|0^{HUM}\rangle$ and $|1\rangle$ itself. Therefore, the energy of the excited states does not constitute a criterion for approaching the exact Hamiltonian eigen-state.]

For atomic and molecular excited states, the highest bound energy is $E[f_1]=0$ and the highest $\langle\phi_1^+|H|f_1\rangle = -E_0>0$. For these systems the largest of the two conditional maxima of $\langle 1|f^{I,II}\rangle^2$, $(1-\delta/\Delta)\langle 1|\phi_1^+\rangle^2$, equals $E_1/E[\phi_1^+]\langle 1|\phi_1^+\rangle^2$. Label by $|f^{I\_best}\rangle$ the corresponding function, for which $E[\phi_1^+] < E[f^{I\_best}] \equiv E_1 \leq E[1^{HUM}]$, and which is orthogonal to $|0^{HUM}\rangle$: $|f^{I\_best}\rangle$ can be expressed in terms of $|1\rangle$ (supposedly known in the present analysis), of $|0^{HUM}\rangle$, and of any known $|\phi\rangle$, used to produce $|f_1\rangle$ orthogonal to both $|1\rangle$ and $|0^{HUM}\rangle$, if $|\phi\rangle$ can yield $\langle\phi_1^+|H|f_1\rangle = 0$ and $E[f_1]=0$: Since

$$|f_1\rangle = \frac{|\phi\rangle\left[1-\langle 1|0^{HUM}\rangle^2\right] - |0^{HUM}\rangle\left[\langle\phi|0^{HUM}\rangle - \langle 1|0^{HUM}\rangle\langle 1|\phi\rangle\right] - |1\rangle\left[\langle 1|\phi\rangle - \langle 1|0^{HUM}\rangle\langle\phi|0^{HUM}\rangle\right]}{\sqrt{1-\langle 1|0^{HUM}\rangle^2}\sqrt{1-\langle 1|0^{HUM}\rangle^2 - \langle\phi|0^{HUM}\rangle^2 - \langle 1|\phi\rangle^2 + 2\langle\phi|0^{HUM}\rangle\langle 1|0^{HUM}\rangle\langle 1|\phi\rangle}} , \qquad (7)$$

it is straightforward to express and solve $\{\langle\phi_1^+|H|f_1\rangle = 0$ and $E[f_1]=0\}$ for $\langle\phi|0^{HUM}\rangle$ and $\langle 1|\phi\rangle$. It turns out that $|\phi\rangle$ must satisfy $\{E_1 (E[0^{HUM}] - E_1 \langle 1|0^{HUM}\rangle) (E[0^{HUM}] E[\phi] - \langle\phi|H|0^{HUM}\rangle^2) \} > 0$, defining those functions $|\phi\rangle$ that can yield $|f_1\rangle = |p\rangle$ (along with $|\phi_1^+\rangle=|m\rangle$, eigenvectors in their planar subspace). Substituting $\langle\phi|0^{HUM}\rangle$ and $\langle 1|\phi\rangle$ back to Eq. 7, $|p\rangle$ is obtained, that yields, by Eq. 5, $|f^{I\_best}\rangle = (E_1/E[\phi_1^+])^{1/2} |\phi_1^+\rangle - (1-E_1/E[\phi_1^+])^{1/2} |p\rangle$, which has the largest projection on $|1\rangle$: $(E_1/E[\phi_1^+])^{1/2} |\langle 1|\phi_1^+\rangle|$ and energy equal to $E_1$, while orthogonal to $|0^{HUM}\rangle$.

It should be noted that minimizing directly $E[1^{HUM}]$, which worsens $|0^{HUM}\rangle$, simply approaches some $|f^{I,II}\rangle$ orthogonal to the new (worse) $|0^{HUM}\rangle$, differing significantly from $|1\rangle$, although it approaches $E[f^{I,II}] \equiv E_1$.

### (iii) $|1^{HUM}\rangle$ Cannot Approach $|f^{I\_best}\rangle$ Because $E[f^{I\_best}] \equiv E_1 < E[1^{HUM}]$, If $|0^{HUM}\rangle \neq |0\rangle$

Since $E[\phi_1^+] < E_1$, the projection $\langle 1|f^{I\_best}\rangle^2 = (E_1/E[\phi_1^+])\langle 1|\phi_1^+\rangle^2 < \langle 1|\phi_1^+\rangle^2$. But since, by HUM theorem, $E[1^{HUM}] > E_1 \equiv E[f^{I\_best}]$, we conclude that $|1^{HUM}\rangle$ cannot approach $|f^{I\_best}\rangle$. Therefore, $\langle 1|1^{HUM}\rangle^2 < \langle 1|f^{I\_best}\rangle^2 = (E_1/E[\phi_1^+])\langle 1|\phi_1^+\rangle^2 < \langle 1|\phi_1^+\rangle^2$. Substituting $|\phi_1^+\rangle$ and $E[\phi_1^+]$ from Eqs. (3, 4) we obtain Eq. 1, the main restriction on the quality of $|1^{HUM}\rangle$. As a consequence:

*If $|0^{HUM}\rangle$ has good quality, then $|1^{HUM}\rangle$ has lower quality*

Indeed, supposing that $|1^{HUM}\rangle \approx |f^{I\_best}\rangle$, then, by writing $\langle 1|0^{HUM}\rangle^2 = 1 - \langle 0|0^{HUM}\rangle^2 - R^2$, it is seen from Eq. 1 that $\langle 1|1^{HUM}\rangle^2 < \langle 0|0^{HUM}\rangle^2$ for a wide range of $\langle 0|0^{HUM}\rangle^2$ and $R^2$. For example, in He $^1S$, where the contribution of $R$ to the energy expansion is written as $E[0^{HUM}] = \langle 0|0^{HUM}\rangle^2 E_0 + \langle 1|0^{HUM}\rangle^2 E_1 + R^2 E_R$, $(-E_1 < E_R < 0)$, by setting indicatively $E_R = -1.7$ $E_h$, then, around the point of the largest difference of $(\langle 0|0^{HUM}\rangle^2 - \langle 1|1^{HUM}\rangle^2)$, i.e. around $(|\langle 0|0^{HUM}\rangle| \approx 0.75, R = 0)$ for a radius $|R| \approx 0.2$, it is seen that $|1^{HUM}\rangle$ has lower quality than $|0^{HUM}\rangle$, because the difference is positive. In this example, as shown in Fig. 1, their relative difference does not exceed ~10% (provided that $|1^{HUM}\rangle \approx |f^{I\_best}\rangle$). If $|1^{HUM}\rangle \neq |f^{I\_best}\rangle$, the difference is larger, since $\langle 1|1^{HUM}\rangle^2$ is smaller than $\langle 1|f^{I\_best}\rangle^2$. The example of the figure refers to He $^1S$, $|0^{HUM}\rangle=1s^2$, $|1^{HUM}\rangle=1s2s$, where the experimental energies have been used (in a.u.), $E_0= -2.9037$, $E_1= -2.146$, $E_2= -2.061$, $E_3= -2.033$, $E_4= -2.021$, [3] both to create the graph, and to estimate indicatively $E_R = -1.7$ $E_h$.

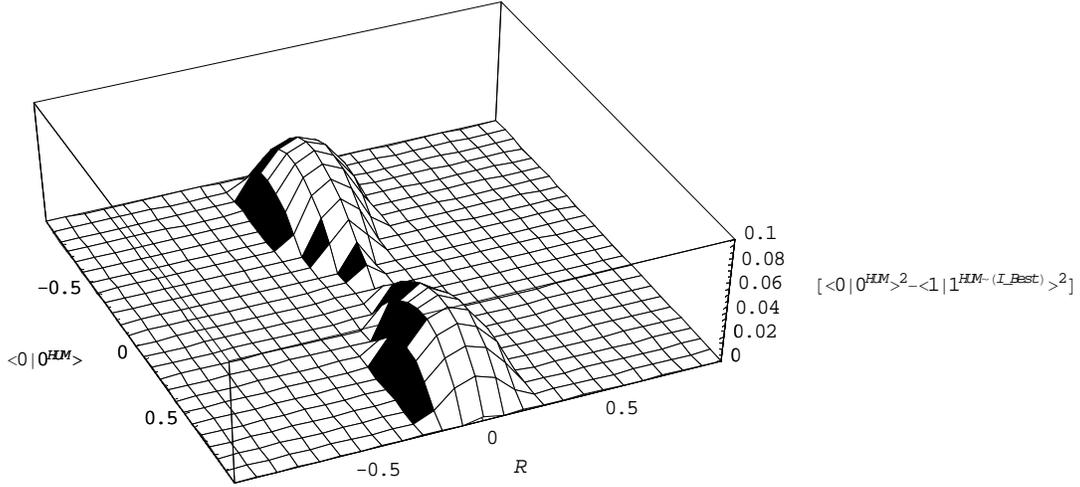

FIGURE 1. The difference of $\{<0|0^{HUM}>^2 - <1|1^{HUM}>^2\}$, provided that $|1^{HUM}>$ is close to $|f^{1\_best}>$, which is the function that: (i) is orthogonal to $|0^{HUM}>$, (ii) has the largest projection on $|1>$, while (iii) it has the energy of $E_1$. The $R$ - axis represents the contribution of all higher than $|1>$ terms in an expansion of the normalized $|0^{HUM}>$ in the exact eigenvectors, i.e. $<0|0^{HUM}>^2 + <1|0^{HUM}>^2 + R^2 = 1$.

## III. CONCLUSION

The HUM theorem, by itself, restricts the higher roots to lie higher than the exact eigen-states, whereas the best approximation of $|1>$ must lie lower than the exact. Thus, the HUM higher roots have some deficiencies (cf. Eq. 1), which prevent them from approaching the exact eigen-states - if the lowest root is not very accurate. The restrictions vanish as $|0^{HUM}>$ tends to $|0>$. For small systems, for which $|0^{HUM}>$ can be obtained accurately enough, the above restrictions on $|1^{HUM}>$, that prevent it from reaching $|1>$, are immaterial, although, of course the quality of the higher roots always remains inferior to that of $|0^{HUM}>$. For larger systems, however, in which $|0^{HUM}>$ is less accurate, the restrictions become significant, rendering $|1^{HUM}>$ unreliable. In such cases (in non-degenerate cases in a common symmetry type), the variational functionals for excited states can be used, which approach $|1>$, $|2>$, …, without any restriction, and without demanding use of very accurate approximants of lower lying states (cf. arXiv:0801.3673).

## ACKNOWLEDGMENTS

The present work was co-funded by EP04111 ENTEP-2004 and by EΔ03968 PENED-2003 grant under measure 8.3 of O.P "Competitiveness" as follows: European Social Fund (75% of public funds), Greek Ministry of Development (GSRT) (25% of public funds), and private funds (10%).